\documentclass [12pt, a4paper]{article}
\usepackage{amssymb}
\begin{document}

\title{A nonparametric CUSUM control chart based on the Mann-Whitney statistic}

\author{
Dabuxilatu Wang\footnote{Corresponding author.
E-mail:\texttt{wangdabu@gzhu.edu.cn} }\quad\quad Qiang Xiong
\\
Department of Statistics, School of Economics \\
and Statistics, Guangzhou University, No. 230 WaiHuan XiLu ,\\
Higher Education Mega Center, Guangzhou,  510006, P.R.China \\
}
\date{}

\maketitle

\begin{abstract}
This article aims to consider a new univariate nonparametric
cumulative sum (CUSUM) control chart for small shift of location
based on both change-point model and Mann-Whitney statistic. Some
comparisons on the performances of the proposed chart with other
charts as well as the properties of the test statistic are
presented. Simulations indicate that the proposed chart is
sensitive in detection of  the small mean shifts of the process by
a high intensive accumulation of sample information when the
underlying variable is completely distribution-free.

\noindent {\bf keywords}: {\em   Change-point, Mann-Whitney
statistic, cumulative sum chart } \noindent{\bf Mathematics
Subject Classification (2000)} 62p30
\end{abstract}

\section{Introduction}
Statistical process control (SPC) has been applied widely for
monitoring various industrial manufacturing processes, service
processes and some special behavior processes (see. Wetherill B.
 et al. \cite{ba}, Montgomery \cite{mo}), in which control charts
are the most widely used to detect changes in production process.
In conventional SPC, the monitored process variable  is assumed to
be modelled by the normal distribution, and based on such
underlying distribution assumption, Shewhart chart,  EWMA
(exponential weighted moving average) chart and CUSUM (cumulative
sum ) chart for variables data, p-chart or c-chart for attributes
data, had been proposed (see. Wetherill B. et al. \cite{ba}). It
is well recognized however that in many applications the
underlying process distribution is not known (e.g. in phase I of
SPC) sufficiently to assume normality (or any other parametric
distribution), so that statistical properties of commonly used
charts, designed to perform best under the assumed distribution,
could be highly affected. In situation like this, development and
application of control charts that do not depend on any specific
parametric distributional assumptions, seem highly desired.  There
are literatures considering distribution-free  charts or
nonparametric charts for this purpose,  an extensive overview on
univariate nonparametric control charts was presented by
Chakraborti \cite{ch}. While the average charts are probably the
most widely used in detection of large mean shift because of their
simplicity, CUSUM procedures are quite appropriate in view of the
sequential nature of the process problem and more sensitive to
small shift in process mean. Based on the within group Wilcoxon
signed-rank statistic, Bakir and Reynold \cite{ba1} proposed a
nonparametric CUSUM chart to track the shift of a location
parameter $\mu$ from an in-control known value $\mu_0$. McDonald
\cite{mc} established some nonparametric CUSUM chart for detecting
the process mean shifts by using sequential rank test. Using
cross-sectional antiranks of the measurement as well as the order
information of the sample, Qiu \cite{qi}\cite{qi1} designed some
nonparametric multivariate CUSUM control charts for detecting
process variability and process mean shifts, respectively.
Recently, in the case of having historical in-control data, Zhou
et al. \cite{zh} established a nonparametric EWMA control chart by
using the Mann-Whitney statistic, which performed a lower
sensitivity than that of change point charts (see. Hawkins et
al.\cite{ha}) having known normal distribution, and  a higher
sensitivity and robustness than that of change point charts (see.
Hawkins et al.\cite{ha})  under non-normal distribution or
distribution-free conditions. Chakraborti \cite{ch1} also proposed
a nonparametric average control chart-MW chart based on the
Mann-Whitney statistic under condition of available reference
sample from the in control process by a phase I analysis, which
has a high sensitivity in detection of the large process mean
shifts in a completely distribution-free case. Das \cite{da}
presented a note on efficiency of nonparametric control chart for
monitoring process variability by using the rank-sum statistic of
Ansari and Bradley. Yang and Cheng \cite{ya} proposed a
nonparametric CUSUM mean chart based on the total number of
univariate data exceeding the in-control mean which is already
known or estimated by the available in control reference sample.

It is well known that the Mann-Whitney statistic is equivalent to
the Wilcoxon rank-sum statistic.  Though the rank based CUSUM
charts have been proposed,  there is no report on a nonparametric
CUSUM chart based on the Mann-Whitney statistic. In this research,
we design such CUSUM chart under the condition of having available
in control reference sample like the cases of Zhou et
al.\cite{zh}, Chakraborti \cite{ch1} and Yang and Cheng \cite{ya},
and a finite sequence of future observations with considering the
concept of change-point for detection of small shifts of the
process mean (the location parameter). Some performance
comparisons with other change point chart and nonparametric charts
are considered. The rest of the paper is organized as follow. In
Section 2, the concept of change-point and the Mann-Whitney
statistic have been recalled. In Section 3, the test statistic of
a nonparametric CUSUM chart based on the Mann-Whitney statistic is
proposed, and the relevant control limits and control rule are
designed. In Section 4, some performance comparisons of the
designed chart with other charts are given.

\section{Change-point and the Mann-Whitney statistic}
The traditional change-point model can be illustrated as follows.
Let $\{X_i,i=1,2,\cdots\}$ be a sequence of independent random
variables, and
$$
X_i\sim\left\{\begin{array}{ll} F(x;\mu_0,\sigma^2_0), &
i=1,2,\cdots,\tau;\\
F(x;\mu_1,\sigma^2_1), & i=\tau +1,\tau +2,\cdots
\end{array}\right.
$$
where $F$ stands for a continuous distribution function with
unknown types, $\mu_0,\mu_1$ stand for the process mean and
$\sigma^2_0,\sigma^2_1$ the process variance. If it holds
$\mu_0\ne\mu_1$ or $\sigma^2_0\ne\sigma^2_1$, then $\tau$ is said
to be a change-point. Detecting and finding out a change-point for
random process is an important issue since with which we can
predict properly the potential change in the observed process. The
detection of the mean shifts or variance change in process control
is similar to the detection of change-point in a process. Like
Zhou et al. \cite{zh} we also consider a change-point model based
on sequence of finite independent random variables $\{X_i\}$,
where $i=1,2,\cdots,l.$

In the nonparametric data analysis, a sort of change-point
detection method is the well-known Mann-Whitney two-sample test
(see. Mann and Whitney \cite{ma}), which is applied for inferring
whether  differences exist between the distributions of two
populations through two independent group samples. For any
$1\leqslant t<l$, the Mann-Whitney statistic is defined as
$$
MW_{t,l}=\sum_{i=1}^t\sum_{j=t+1}^lI(x_j<x_i)=\sum_{j=t+1}^l
I(x_j<x_1)+I(x_j<x_2)+\cdots +I(x_j<x_t),
$$
where
$$
I(x_j<x_i)=\left\{\begin{array}{ll} 1, & x_j<x_i,\\
0,  & x_j\geqslant x_i.
\end{array}\right.
$$
If $\mu_0=\mu_1,\sigma^2_0=\sigma^2_1$, then it is said that the
process is in-control state. In this paper, we only consider the
variation on the location parameter and let
$\sigma^2_0=\sigma^2_1$.
 It is verified that (see. Mann and Whitney \cite{ma}, Zhou et
al.\cite{zh}), under in-control state, the expectation and
variance of $MW_{t,l}$ can be obtained as
$$
E_0 (MW_{t,l})=\frac{t(l-t)}{2}, Var_0
(MW_{t,l})=\frac{t(l-t)(l+1)}{12}.
$$
The standardized Mann-Whitney statistic $SMW_{t,l}$ is defined by
$$
SMW_{t,l}=\frac{MW_{t,l}-E_0 (MW_{t,l})}{\sqrt{Var_0 (MW_{t,l})}}.
$$
Based on the proof of Mann-Whitney (see. Mann and Whitney
\cite{ma}),  when the process is in-control state, the
distribution of $SMW_{t,l}$ is symmetric about zero for each $t$,
and large values of $SMW_{t,l}$ indicate a negative mean shift,
whereas small values indicate a positive shift \cite{zh}. As
explained in Zhou et al. \cite{zh}, a test statistic for detection
of change-point about the mean (i.e. the hypotheses
$H_0:\mu_0=\mu_1$) is proposed by Pettitt \cite{pe} as
$$
T_l=\max_{1\leqslant t\leqslant l-1}|SMW_{t,l}|.
$$
If $T_l$ exceeds some critical value $h_l$, then we conclude that
there is a shift in the mean. Otherwise, we conclude that there is
no sufficient evidence of a shift. To find a suitable critical
values $h_l$, we can use the limiting distribution of $T_l$ given
by Pettitt \cite{pe} to make an approximation.

Note that for each $t$, $SMW_{t,l}$ can be modeled approximately
by standard normal distribution $N(0,1)$ when $l$ is large.

\section{A nonparametric CUSUM chart}
For detecting  a small mean shift occurred at the change-point of
the process as soon as possible, CUSUM control chart is usually
recommended. Assuming that the occurred change-point indicates an
upward mean shift, i.e., $\mu_1>\mu_0$, then at the change-point,
the expectation of the Mann-Whitney statistic becomes larger and
the sample mean around the change-point might occurs a small
upward shift. In such case, it is hard to find shift quickly by
using the statistic $T_l$,  because it only depends on the finite
individual observations without considering all historical sample
information. Therefore, we desire to construct an upper side
nonparametric CUSUM control chart based on the standardized
Mann-Whitney statistic $SMW_{t,l}$.

Zhou et al. \cite{zh} have done an improvement on statistic $T_l$
and proposed the $SMW$ chart, which has a test statistic
$T_{m,n}=\max_{m\leqslant t<m+n}|SMW_{t,(m+n)}|$. Furthermore, an
EWMA chart has been given based on $SMW_{t,(m+n)}$, where $m$ is
the number of in-control historical individual observations, $n$
is the number of the future observations, and $m+n$ is the number
of the total observations.  Noting that here the CUSUM chart has
not been considered. Though both CUSUM  chart  and EWMA chart
performed quite well in the detection of the small shifts in mean
(see. Lucas and Saccucci \cite{lu}), CUSUM chart is usually
slightly more sensitive than EWMA chart when the average in
control run length become large (see. Srivastava and Wu
\cite{sr}). Under the same condition of the $SMW$ chart, we
propose the test statistic for our upper side CUSUM chart as
$$
S_j(m,n)=\max\{0,S_{j-1}(m,n)+SMW_{j,(m+n)}-k\}, j=m-m_0,m-m_0
+1,\cdots,m-m_0+n-1,
$$
where $0\leqslant m_0\leqslant m$, $S_{m-m_0-1}(m,n)=0$,
$k=\frac{\vartriangle}{2}$ is the reference value, $\vartriangle$
is the shift size to be detected. Here we assume $k=\frac{1}{2}$.
Set
$$
S_{max}(m,n)=\max_{m-m_0\leqslant j\leqslant
m+n-1}|S_j(m,n)|=\max_{m-m_0\leqslant j\leqslant m+n-1} S_j(m,n).
$$
Our control rule is as follow,\\
(1)After the $n$th future sample is monitored, compute
$S_{max}(m,n)$.\\
(2)Let $h_{m,n}$ be the decision value, which is chosen to obtain
the given in-control average run length. If $S_{max}(m,n)\leqslant
h_{m,n}$, we conclude that there is no evidence of a shift and
continue to monitor the $(n+1)$st future sample. If
$S_{max}(m,n)\geqslant h_{m,n}$, then an out-of-control signal is
triggered.

Noting that, in the case of SMW chart (see. Zhou et al.
\cite{zh}), we need to calculate the maximum values of
$SMW_{t,(m+n)}$ for each $t$, whereas for our CUSUM chart, we
calculate the maximum values of the cumulative sum $S_j(m,n)$.

For the given type one error $\alpha$, the decision value
$h_{m,n}$ can be obtained by solving following equations
$$
Pr(S_{max}(m,n)>h_{m,n}(\alpha)|S_{max}(m,i)\leqslant
h_{m,i}(\alpha), 1\leqslant i<n)=\alpha, n>1,
$$
$$
Pr(S_{max}(m,1)> h_{m,1}(\alpha))=\alpha.
$$

Due to the intricacy of this conditional probability, it seems to
be impossible to solve it analytically. Therefore, similar to
\cite{zh}, we use one million sequences of length 500 which come
from the standard normal distribution to estimate them. The
historical sample size is assumed to be larger than 10. Table 1
shows the control limit of CUSUM chart for $\alpha$ values of
0.01, 0.005, 0.0027, and 0.002, corresponding to in control ARLs
of 100, 200, 370, 500, for $m$=10  and  50, $m_0$=4, and $n$
values in the range 1-490. As shown in Table 1, $h_{m,n}(\alpha)$
increase initially, but then stabilizes. We can obtain the optimal
decision value using such approach of  estimation. Compare with
the decision values $h_{m,n}(\alpha)$ shown in Table 1 of Zhou et
al. \cite{zh} for their EWMA chart, the decision values of CUSUM
chart seem slightly smaller. Similar to Zhou et al.\cite{zh}, it
is not difficult to present an illustrative example to introduce
the implementation of our proposed CUSUM chart, we omit it here.

The equivalence between the Wilcoxon rank-sum statistic $W_{t,l}$
and the Mann-Whitney statistic $MW_{t,l}$ is shown with the
equality $MW_{t,l}=W_{t,l}-\frac{t(t+1)}{2}$, where
$W_{t,l}=\sum_{i=1}^t R_i$, and $R_i$ denotes the rank of the
$i$th observation $x_i$ in the total $l$ observations. We may use
this equivalence to reduce the computational complexity of the
Mann-Whitney statistic.
\begin{table}
\begin{footnotesize}
Table 1\\
The decision values $h_(m,n)(\alpha)$ of CUSUM control chart, $k=\frac{1}{2}$\\
\begin{tabular}{|l|l l l l l l l l|}
\hline
     & m=10 &   &   &   & m=50 &  &  & \\
\hline
     & ARL(0)   &   &   &   &   &  &  & \\
\hline
   n  &  100   & 200&370&500&100&200&370&500 \\
\hline
1 &1.120&1.185&1.184&1.114&1.182&1.192&1.235&1.254\\
3 &1.263&1.329&1.216&1.412&1.346&1.395&1.459&1.502\\
5 &1.462&1.589&1.668&1.693&1.554&1.660&1.748&1.827\\
7 &1.611&1.752&1.865&1.914&1.696&1.847&1.974&2.038\\
9 &1.689&1.845&1.972&2.021&1.774&1.945&2.082&2.150\\
11 &1.744&1.916&2.050&2.119&1.833&2.023&2.160&2.238\\
13 &1.798&1.988&2.129&2.167&1.887&2.082&2.238&2.316\\
15 &1.836&2.027&2.178&2.246&1.996&2.121&2.297&2.375\\
17 &1.873&2.076&2.227&2.295&1.950&2.160&2.336&2.484\\
19 &1.897&2.090&2.261&2.334&1.970&2.180&2.360&2.453\\
22 &1.924&2.129&2.295&2.393&1.989&2.198&2.394&2.472\\
26 &1.951&2.170&2.349&2.432&1.999&2.238&2.434&2.522\\
30 &1.985&2.191&2.383&2.481&2.038&2.192&2.453&2.551\\
35 &2,065&2.237&2.422&2.510&2.043&2.297&2.483&2.581\\
40 &2.026&2.268&2.452&2.549&2.063&2.367&2.502&2.60\\
50 &2.049&2.292&2.496&2.588&2.078&2.327&2.522&2.620\\
60 &2.071&2.317&2.530&2.628&2.097&2.346&2.552&2.649\\
70 &2.083&2.333&2.546&2.638&2.107&2.356&2.571&2.669\\
80 &2.091&1.341&2.559&2.657&2.112&2.366&2.591&2.688\\
90 &2.095&2.356&2.569&2.667&2.117&2.386&2.60&2.698\\
115 &2.106&2.372&2.589&2.686&2.122&2.386&2.605&2.708\\
140 &2.118&2.380&2.60&2.6906&2.127&2.390&2.610&2.718\\
165 &2.152&2.388&2.608&2.706&2.129&2.395&2.655&2.727\\
190 &2.136&2.405&2.628&2.726&2.142&2.415&2.63o&2.742\\
240 &2.134&2.413&2.633&2.735&2.147&2.420&2.635&2.747\\
290 &  &2.421&2.643&2.745&  &2.425&2.620&2.752\\
390 &  &    &2.658&2.760&   &    &2.659&2.767\\
490 &   &    &   &2.795&     &    &     &2.800\\
\hline
\end{tabular}
\end{footnotesize}
\end{table}

\section{A comparison between control charts}
In this section, we present a simple performance comparison shown
in Table 2 between our CUSUM chart and the change-point chart
proposed by Hawkins et al. \cite{ha} based on the available data
from Zhou et al. \cite{zh}.  Also we consider some comparisons of
the characteristics of our test statistic to other nonparametric
CUSUM statistic.

\subsection{A comparison with change-point chart}

In Table 2, $\delta$ denotes the coefficient of standard deviation
for measuring the shift size, $\tau$ denotes the change-point.
\begin{table}
\begin{footnotesize}
Table 2\\
The ARL comparisons between CUSUM chart and C-PC (Change-Point Chart)\\
for N(0,1) data and m=10, $\alpha$=0.005\\
\begin{tabular}{|l|l|l|l|l|l|l|l|l|}
\hline
     & $\tau$=10 &   & $\tau$=50 &   & $\tau$=100 &  & $\tau$=250 & \\
\hline
 $\delta$ & CUSUM&C-PC&CUSUM&C-PC&CUSUM&C-PC&CUSUM & C-PC \\
\hline
0.00&200.0&200.0&200.0&200.0&200.0&200.0&200.0&200.0\\
0.025 &173.1&187.5&123.4&155.5&96.4&130.7&70.8&100.3\\
0.050 &134.8&159.8&39.9&66.4&26.8&41.1&21.9&31.1\\
0.75 &92.1&113.5&13.9&22.8&11.9&17.0&10.9&15.1\\
1.00 &52.80&65.5&8.7&11.6&8.2&9.9&8.2&9.2\\
1.25 &26.6&30.6&6.9&7.5&6.7&6.7&6.9&6.4\\
1.50 &14.2&15.0&6.3&5.4&5.9&5.0&5.8&4.8\\
1.75 &8.2&8.8&5.2&4.2&5.3&3.9&5.1&3.7\\
2.00 &5.8&6.3&4.3&3.4&4.3&3.2&4.2&3.0\\
2.25 &5.4&5.0&3.9&2.8&3.9&2.6&3.9&2.6\\
2.50 &4.2&4.2&3.4&2.4&3.5&2.3&3.5&2.2\\
2.75 &3.8&3.6&3.1&2.1&3.1&2.0&2.9&1.9\\
3.00 &3.5&3.1&2.8&1.8&2.9&1.8&2.6&1.7\\
\hline
\end{tabular}
\end{footnotesize}
\end{table}
We can found that\\
(1) As the increasing of the future  observed in-control data,
both charts become more sensitive to the shift as the new
observation updates the information already known.\\
(2) For detecting a relatively large shift in mean, the
change-point chart proposed by Hawkins et al. \cite{ha} is more sensitive than our CUSUM chart.\\
(3) Our CUSUM chart is faster than the change-point chart proposed
by Hawkins et al. \cite{ha} for detecting a small mean shift.

Similar to Zhou et al. \cite{zh}, we may make other performance
comparisons of our CUSUM chart to other control charts having
known underlying distribution, we omit it here.

\subsection{A comparison with other nonparametric CUSUM charts}

In the following, we also mention a rough comparison between our
CUSUM chart and other nonparametric CUSUM charts  proposed by
Bakir and Reynolds \cite{ba1},  McDonald \cite{mc} and Yang and
Cheng \cite{ya} based on the characteristics of the test
statistic.

Bakir and Reynolds \cite{ba1} consider a sequence of observations
$\{x_{ij},i=1,2,\cdots;j=1,\cdots,g\}$, each of size $g=4$ or $5$,
and define a within group Wilcoxon signed rank sum
$SR_i=\sum_{j=1}^g sign(x_{ij})R_{ij}$ for each observation, where
$R_{ij}$ denotes the rank of $|x_{ij}|$ in
$\{|x_{i1}|,\cdots,|x_{ig}|\}$,  and based on which they propose
their CUSUM statistic as
$$
\sum_{i=1}^n (SR_i -k) -\min_{0\leqslant m\leqslant n}\sum_{i=1}^m
(SR_i -k),
$$
or
$$
\max_{0\leqslant m\leqslant n}\sum_{i=1}^m (SR_i +k)-\sum_{i=1}^n
(SR_i +k).
$$
From the structure of the statistic, we are not able to compare it
straightforwardly with our CUSUM statistic because the rank in
Mann-Whitney is based on the comparison of the total samples,
whereas the rank in former statistic  is only based on comparison
within group sample with its absolute values. If we restrict a
sequence of group sample of size $g$ to finite $l$ times
observations, then we may view them as a finite sequence of
$l\times g$ independent random variables, and we may assume the
former $m$ variables of the sequence are in-control state, so that
we can obtain our standardized Mann-Whitney statistic with a
reference in-control data. Noting that the rank involved in our
CUSUM statistic is for total samples and the rank in Bakir and
Reynolds 's CUSUM statistic \cite{ba1} is only the within group
signed rank, the sum of the former ranks is obviously larger than
the sum of the later ranks. Therefore,  for a fixed reference
value $k$ and a decision value $h$, our CUSUM chart is clearly
more sensitive than Bakir and Reynolds's chart for small mean
shift.

For comparing our CUSUM chart with McDonald's  CUSUM chart
\cite{mc}, we note that their sequential rank $R_i$ is defined as
$R_i=1+\sum_{j=1}^{i-1} I(x_j < x_i)$ for the observation $\{x_i,
i=2,\cdots\}$, and the CUSUM  statistic $T_j=\max\{0,  T_{j-1}+U_j
-k \}$, where $U_i=\frac{R_i}{i+1}$. The Mann-Whitney statistic
$MW_{i,n}$ can be written as
$$
MW_{i,n}=(R_{i+1} -1)+ \sum_{s=i+2}^n (I\{x_1 < x_s\}+I\{x_2 <
x_s\}+ \cdots + I\{x_i < x_s\}),
$$
$i=2,\cdots,n-1.$ Therefore, $MW_{i,n}$ is obviously larger than
$R_i$, which leads to that the cumulative  sample information of
our CUSUM statistic is more rich than the cumulative sample
information of $T_i$. So our CUSUM chart based on $MW_{i,n}$ is
more sensitive than the CUSUM chart based on $R_i$ proposed by
McDonald \cite{mc}.

We now consider roughly to compare  our CUSUM chart with Yang and
Cheng's CUSUM chart \cite{ya}.  Noting that their chart is based
on statistic $M_t=\sum_{j=1}^{t} I\{x_j
>\mu\}$ for the observations $\{x_i, i=1, 2, \cdots\}$, where $\mu$ is
the in control mean of the process. Obviously, the Mann-Whitney
statistic $MW_{i,n}$ implies more information than $M_t$,
especially in the case where the unknown in control mean need to
be estimated by the available reference in-control samples. They
only utilize the average information like $\bar{\bar{x}}$ of  the
available reference in-control samples, whereas in our CUSUM case,
the $m$ available reference in control samples are fully utilized
with an individually comparison. So our CUSUM chart is also
relatively  more sensitive than Yang and Cheng's CUSUM chart.

\noindent\textbf{Conclusion remark} The rank based statistical
method is an important well-known nonparametric approach for the
case where the distribution of the underlying variable is
completely unknown, in which the Mann-Whitney statistic is the
popular and powerful one. We establish a sort of nonparametric
CUSUM chart based on the standardized  Mann-Whitney statistic for
the detection of the  small location shifts quickly. Comparisons
indicate that the proposed CUSUM chart is slightly quicker than
the nonparametric charts proposed by Zhou et al. \cite{zh}, Bakir
and Reynolds \cite{ba1}, McDonals \cite{mc} and Yang and Cheng
\cite{ya} in the detection of the small mean shifts. However, the
computation of our CUSUM statistic is somewhat difficult, we need
to solve it with the computer programming. It is also to be
pointed out that the nonparametric control charts is rest on the
theoretical research only, of which the development of the
practical applications is highly desired.


\begin{thebibliography}{99}


\bibitem{ba}Barrie Wetherill G. and Brown D. W., (1991), Statistical process control, Theory and practice,
Chapman and Hall, London, New York, Toyko, Melbourne, Madras,

\bibitem{ba1}Bakir, S.T. and  Reynolds, M.R.JR , (1979), A nonparametric procedure for process control based on within
group ranking,  Technometrics,  21: 175-183.

\bibitem{ch}
Chakraborti S., Van Der Laan P. and  Bakir S.T.,  (2001),
Nonparametric control charts: An  Overview  and  Some Results,
Journal of quality technology, 33 (3): 304-315.

\bibitem{ch1}
Chakraborti S. and Van De Wiel Mark A.,  (2008), A nonparametric
control chart based on the Mann-Whiteney statistic, IMS
collections, beyond parametrics in interdisciplinary research:
festschrift in honor of profesor Pranab K. Sen, Vol 1: 156-172.

\bibitem{da} Das, N., (2008), A note on the efficiency of nonparametric
control chart for mornitoring process variability, Economic
quality control, 23(1): 85-93.

\bibitem{ha} Hawkins, D.M., Qiu, P., Kang, C.W., (2003), A changepoint model for statistical process
control. J Qual. Technol.  35: 355-366.

\bibitem{lu} Lucas,J.M., Saccucci, M.S., (1990), Exponentially weighted
moving average control charts schemes: properties and
enhancements, Technometrics, 32:1-30.

\bibitem{ma} Mann H.B., Whitney D.R., (1947), On the test whether one of
two random variables is stochastically larger than other. Ann.
Math. Stat.  18:50-60.

\bibitem{mc} McDonald, D.,  (1990),   A CUSUM  procedure  based on sequential ranks, Naval research
logistics, 37: 627-646.

\bibitem{mo}Montgomery, D.C., (2004), Introduction to statistical quality control,
5th edition. New York: Wiley.

\bibitem{pe} Pettitt A.N., (1979),  A non-parametric approach to the
change-point problem. Applied Statistics.  28: 126-135.

\bibitem{qi} Qiu, P., Hawkins, D.M.,  (2001), A rank based multivariate CUSUM procedure, Technometrics,
 43: 120-132.

\bibitem{qi1} Qiu, P., Hawkins, D.M., (2003),  A nonparametric multivariate CUSUM procedure for detecting shifts in all
directions, Statistician, 52: 152-164.

\bibitem{sr} Srivastava, M.S., Wu, Y., (1993), Comparison of EWMA, CUSUM
and Shiryayev-Roberts procedures for detecting a shift in the
mean, The annals of statistics,  21(2): 645-670.

\bibitem{ya} Su-fen Yang and Smiley W. Cheng,  (2010), A new nonparametric CUSUM mean chart, Quality and reliability engineering
international, doi: 10.1002/qre.1171.

\bibitem{zh} Zhou Chunguang, Zou Changliang, Zhang Yujuan, Wang Zhaojun. (2009),  Nonparametric control
chart based on change-point model. Statistical Papers. 50:13-28.















\end{thebibliography}
\end{document}